\documentclass[fleqn,twoside]{article}
\usepackage{espcrc2}

\usepackage{graphicx}
\usepackage{amssymb}

\newcommand{\beq}{\begin{equation}}
\newcommand{\eeq}{\end{equation}}
\newcommand{\bqa}{\begin{eqnarray}}
\newcommand{\eqa}{\end{eqnarray}}

\newcommand{\AmS}{{\protect\the\textfont2
  A\kern-.1667em\lower.5ex\hbox{M}\kern-.125emS}}

\begin{document}

\thispagestyle{empty}

\title{
\vspace{-1.8cm}
\hfill \rm \null \hfill
 \hbox{\normalsize SNUTP-03-018} \\
\vspace{+1.3cm}
Comparison of $\overline{\rm Fat7}$ and HYP fat links}

\author{S. Bilson-Thompson and W. Lee 
        \address{School of Physics, 
	  Seoul National University, Seoul 151-747 South Korea}}

\begin{abstract}
We study various methods of constructing fat links based upon the HYP
(by Hasenfratz \& Knechtli) and $\overline{\rm Fat7}$ (by W. Lee)
algorithms.  We present the minimum plaquette distribution for these
fat links. This enables us to determine which algorithm is most
effective at reducing the spread of plaquette values.
  
\vspace{1pc}
\end{abstract}

\maketitle

\section{INTRODUCTION}
Despite its prominent role in the non-perturbative study of QCD,
lattice field theory has long struggled with the discretization errors
inherent in its formulation, with the consequent development of many
improvement schemes. One popular method of improvement has been the
use of ``fat links'', in which each ordinary or ``thin'' link of the
lattice, $U_{\mu}(x)$, is replaced by a linear combinaton of the thin
link and links adjacent to it. Two fattening algorithms which will be
of interest to us in this report are HYP or hypercubic
blocking~\cite{hasen}, and SU(3)-projected Fat7 blocking~\cite{wlee:0}
($\overline{\rm Fat7}$). These two algorithms have been shown to be
perturbatively equivalent at one-loop
level~\cite{wlee:0,leesharpe}. Here we present the results of a
preliminary numerical study to determine if HYP and $\overline{\rm
Fat7}$ significantly differ non-perturbatively.

\section{HYPERCUBIC BLOCKING}
The HYP algorithm is a modification of standard APE
smearing~\cite{APE}. The APE-smeared link $V_{\mu}(x)$ is formed by
adding the sum of staples, weighted by some factor $\alpha$, to the
thin link, 
\beq V_{\mu}(x) = (1-\alpha)\,U_{\mu}(x) + \frac{\alpha}{6}
\sum_{\nu=0}^3 W_{\nu}(x)
\label{eq:fatten}
\eeq
followed by projection back to SU(3) ($W_{\nu}(x)$ refers to the sum
of three-link staples in the positive and negative $\nu$ directions,
with central link parallel to $U_{\mu}(x)$, and $\nu\neq\mu$,). To
localize the smearing within the smallest possible volume, HYP
blocking was defined as being equivalent to three iterations of APE
smearing with the caveat that the staples at each stage could not be
constructed from links which are fattened in the same plane as the
staple itself.  Due to space constraints in this report, we direct the
reader to reference~\cite{hasen} for a more detailed description. \\

Since its introduction, HYP blocking has shown itself to be extremely
effective at reducing taste-symmetry breaking effects, and has gained
widespread attention and use.
  
\section{FAT7 AND $\overline{\rm FAT7}$ BLOCKING}
\label{sec:Fat7}
Fat7 blocking~\cite{Fat7} incorporates not only three-link staples but
also five-link and seven-link staples, making it possible to traverse
all three directions orthogonal to the original thin link, as in HYP
blocking. Unlike HYP, the original proposal for Fat7 blocking did not
incorporate SU(3) projection. $\overline{\rm Fat7}$ is a modification
of Fat7 which incorporates SU(3) projection after fattening
\cite{wlee:0}.

We can use the second covariant derivative~\cite{LePage} operator to
define a fattened link
\beq 
L_{\nu}(\alpha) \cdot U_{\mu}(x) = (1 -
2\alpha)\cdot U_{\mu}(x) + \alpha W_{\nu}(x).  
\eeq 
($W_{\nu}(x)$ defined as in Eq.~(\ref{eq:fatten})). With the parameter
$\alpha$ taking the value one-quarter this operator can be interpreted
as suppressing flavour-changing gluon interactions by vanishing in the
limit as gluon momentum approaches the lattice cut-off. The Fat7 link
can easily be constructed by repeated application of this operator as
\beq 
V_{\mu} = \frac{1}{6}\sum_{{\rm
perm}(\nu,\rho,\lambda)} \!\!\!\!\! L_{\nu}(\alpha)\cdot
\left(L_{\rho}(\alpha)\cdot \left(L_{\lambda}(\alpha)\cdot
U_{\mu}\right) \right) 
\eeq 
where we note explicitly that the sum over permutations of the
directions is taken.
\begin{figure}
\includegraphics[angle=90,width=14pc]{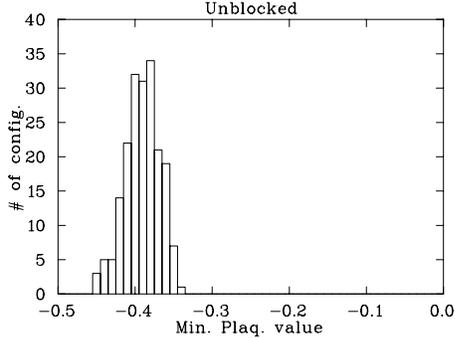} \vspace{-8mm}
\caption{minimum plaquette values for an ensemble of unblocked configurations.}
\label{fig:unblock}
\end{figure}
This repeated use of the same operator means we use a single weighting
parameter $\alpha$ at each stage, whereas HYP uses three parameters
$\alpha_1$, $\alpha_2$, $\alpha_3$. For $\overline{\rm Fat7}$ we may
construct four different combinations of SU(3) projection options
(i.e. projection at the final, initial+final, middle+final, and
initial+middle+final stages).

\section{COMPARISON OF FATTENING APPROACHES}
\begin{figure}[t!]
\includegraphics[angle=90,width=14pc]{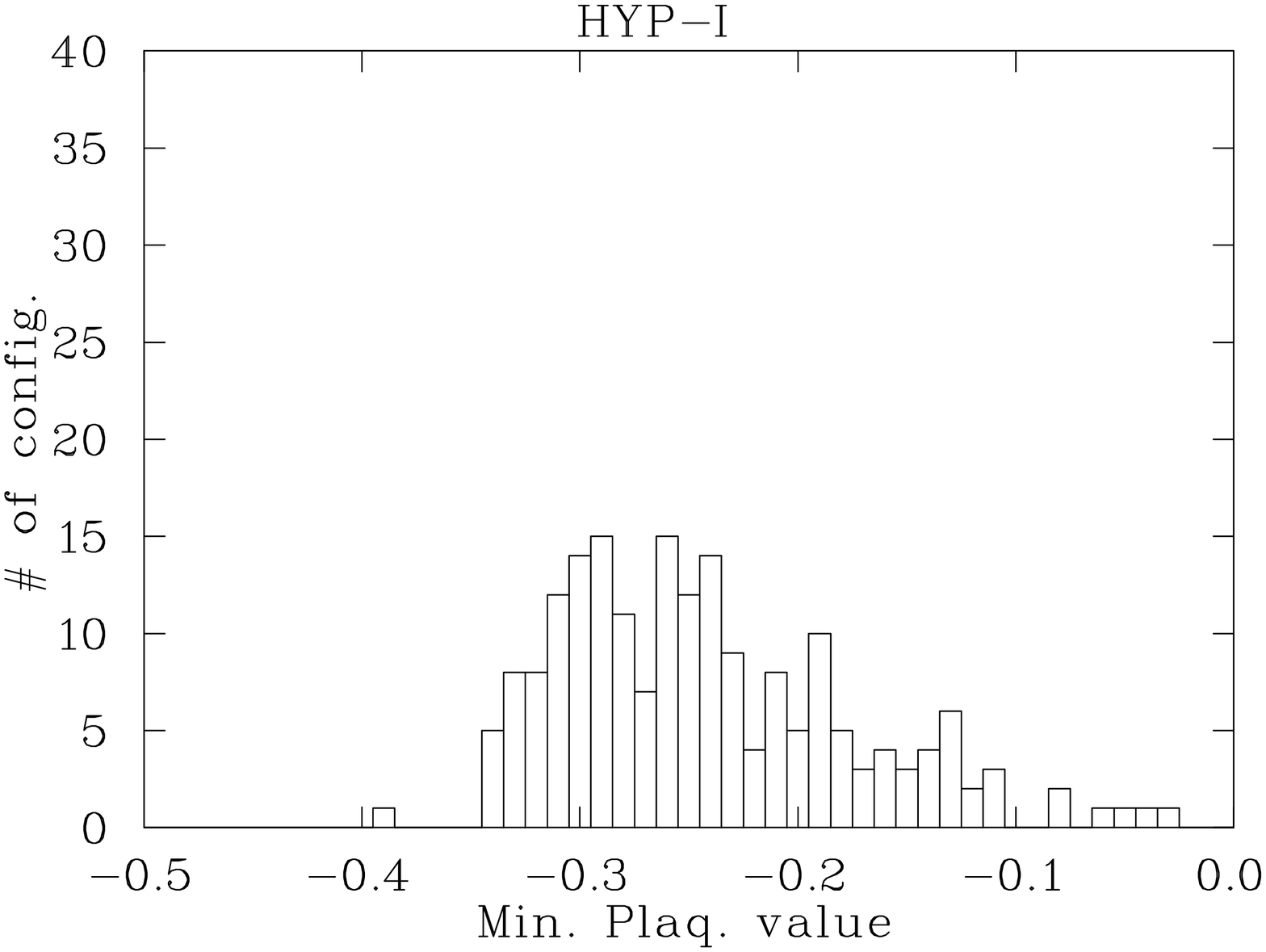} \vspace{-8mm}
\caption{Minimum plaquette values for HYP-I blocked configurations.}
\label{fig:HYP1}
\vspace{2mm}
\includegraphics[angle=90,width=14pc]{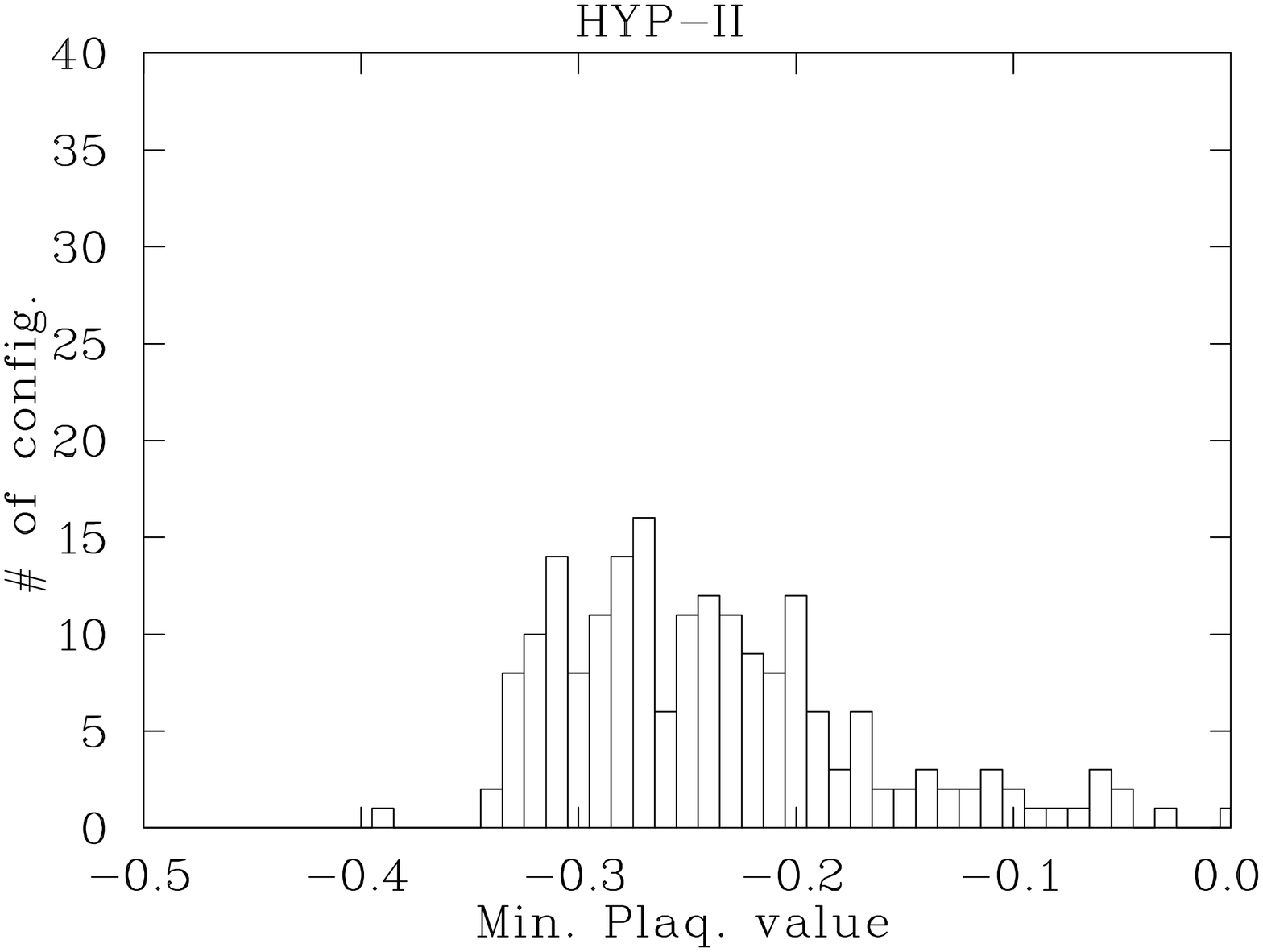} \vspace{-8mm}
\caption{Minimum plaquette values for HYP-II blocked configurations.}
\label{fig:HYP2}
\end{figure}

The process of fattening tends to spread the minimum plaquette values 
of an ensemble of configurations over quite a wide range (compare 
Figs.~\ref{fig:unblock} and~\ref{fig:HYP1}-\ref{fig:HYP2}). In~\cite{hasen}, minimum 
plaquette values were used as a probe of the most severe short-range
link fluctuations, which can lead to taste-symmetry violating effects and 
other problems. We expect that for certain types of measurements it will be 
desirable to reduce link fluctuations by increasing minimum plaquette values, while 
maintaining a narrow spread of minimum plaquette values across our ensemble to reduce 
statistical uncertainty. However it is worth noting that this is a speculative point. 
In this study we wish to determine whether the $\overline{\rm Fat7}$ 
blocking algorithm is capable of producing a narrower spread of minimum plaquette 
values than HYP, making it worthy of a more detailed investigation.

\section{RESULTS}
\begin{figure}
\includegraphics[angle=90,width=14pc]{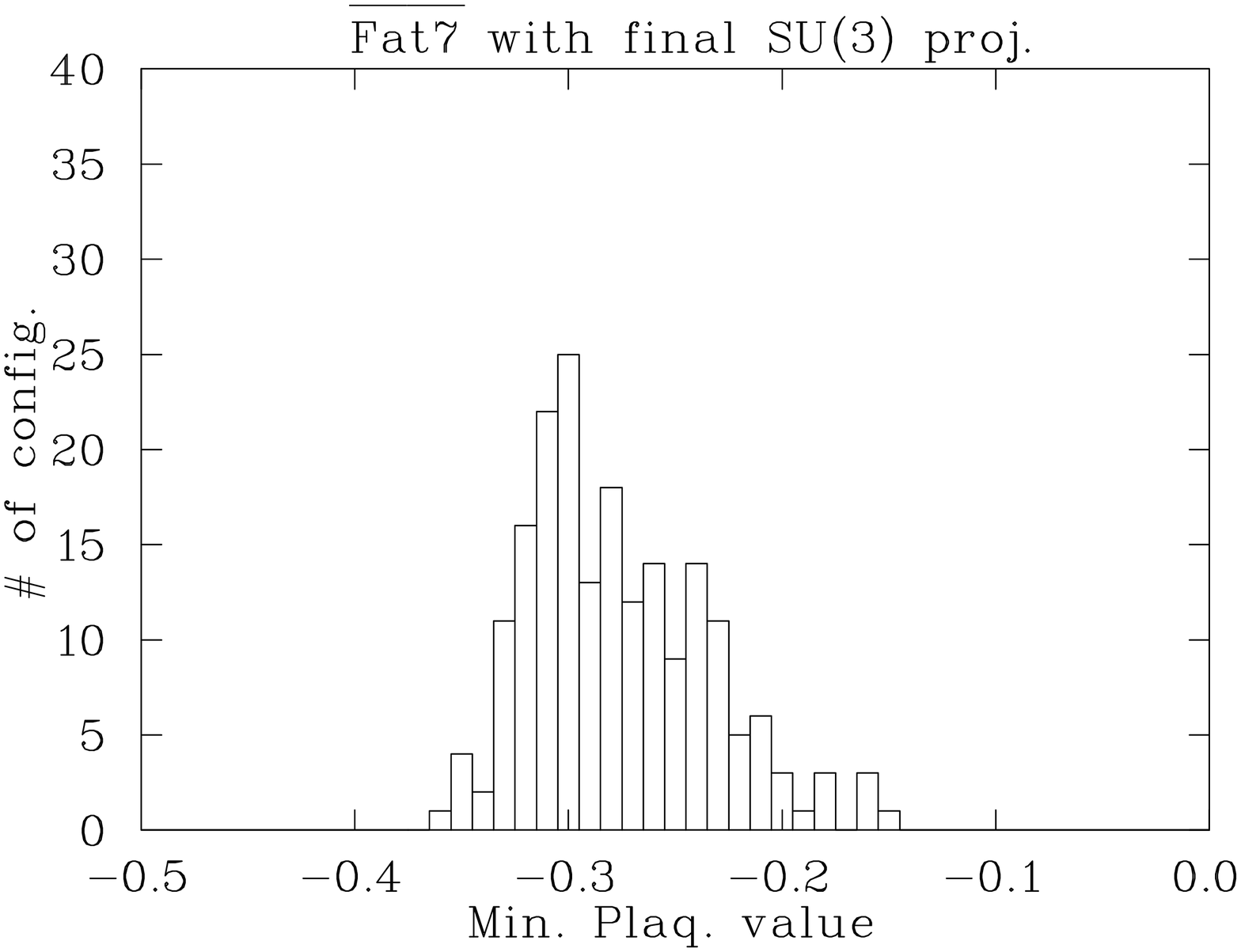} \vspace{-8mm} 
\caption{Minimum plaquette values for $\overline{\rm Fat7}$-blocked configurations, with 
final stage SU(3) projection.} 
\label{fig:Fat7_0}
\vspace{2mm}
\includegraphics[angle=90,width=14pc]{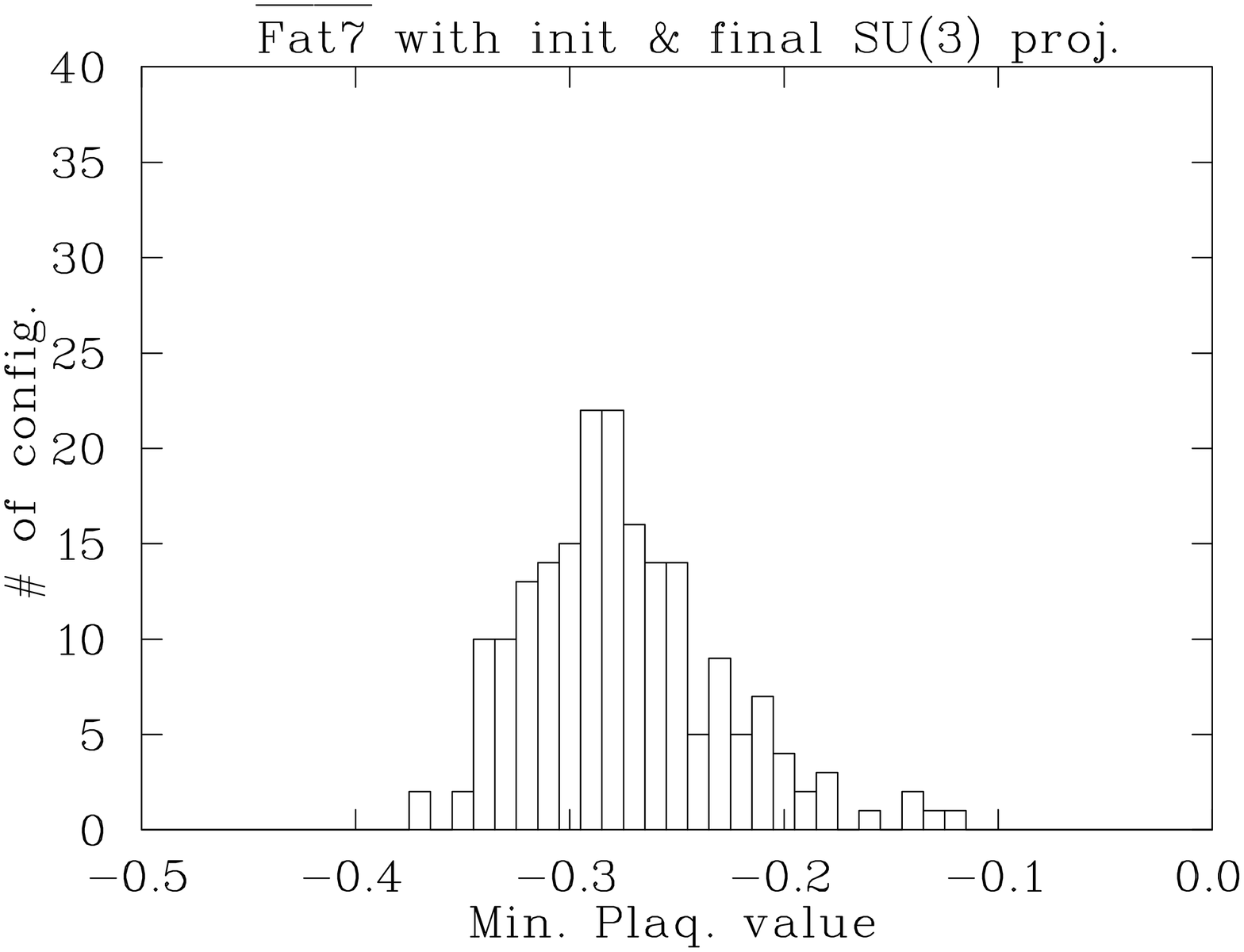} \vspace{-8mm}
\caption{Minimum plaquette values for $\overline{\rm Fat7}$-blocked configurations, with 
initial and final stage SU(3) projection.} 
\label{fig:Fat7_II}
\vspace{2mm}
\includegraphics[angle=90,width=14pc]{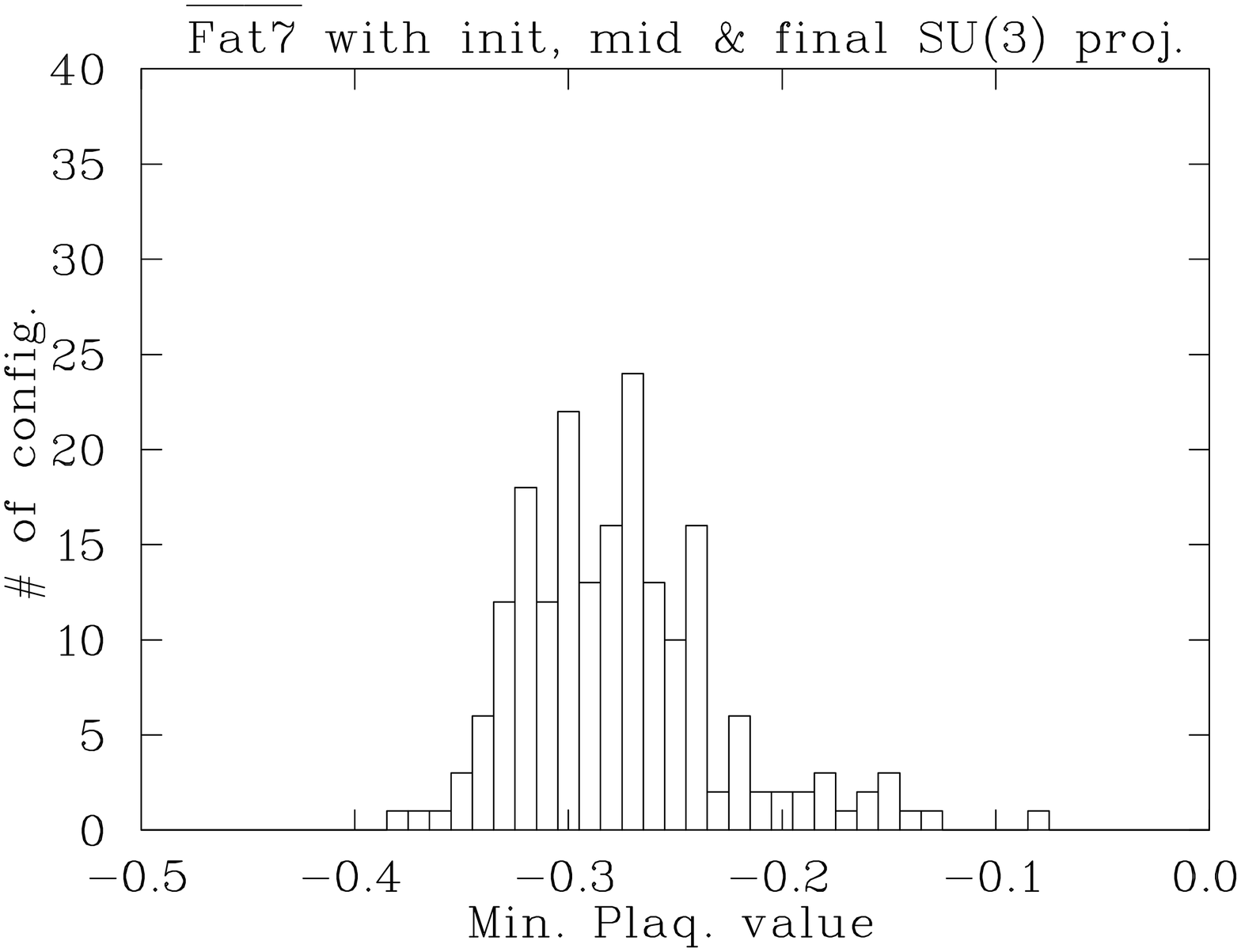} \vspace{-8mm}
\caption{Minimum plaquette values for $\overline{\rm Fat7}$-blocked configurations, with 
initial, middle, and final stage SU(3) projection.}
\label{fig:Fat7_III}
\end{figure}

We present the minimum plaquette values for an ensemble of 194 quenched configurations 
generated on an $8^3 \times 32$ lattice at $\beta = 5.7$. We also present the minimum 
plaquette values for the ensembles obtained after blocking each configuration with
the HYP algorithm (Fig.~\ref{fig:HYP1}-\ref{fig:HYP2}), and $\overline{\rm Fat7}$ with 
$\alpha=0.25$ (Figs.~\ref{fig:Fat7_0}-\ref{fig:Fat7_III}). For HYP we are interested in 
two choices of parameters, those found non-perturbatively 
in~\cite{hasen}, namely $\alpha_1=0.75$, $\alpha_2=0.6$, $\alpha_3=0.3$ (HYP-I), and 
those found perturbatively in~\cite{leesharpe}, namely $\alpha_1=7/8$, $\alpha_2=4/7$, 
$\alpha_3=1/4$ (HYP-II), although we note the caveat that choices of $\alpha_1$ larger 
than 0.75 may tend to destabilize the smearing algorithm~\cite{toomuch}. For the 
$\overline{\rm Fat7}$-blocked configurations we have analysed the four different SU(3) 
projection options described in section~\ref{sec:Fat7}. The $\overline{\rm Fat7}$ 
configurations show less spread than HYP. $\overline{\rm Fat7}$ with projection at the 
final stage, and at all three stages are interesting from an analytical 
basis~\cite{leesharpe}. Although the number of configurations used in this study 
is small, the narrow, well-defined peak obtained after blocking with initial+final 
stage SU(3) projection seems significant enough to also warrant further investigation.

\section{CONCLUSIONS}
In this preliminary work $\overline{\rm Fat7}$ shows signs of producing results which 
are as good as HYP, and possibly better for some calculations. Both HYP and 
$\overline{\rm Fat7}$ can be constructed very efficiently by pre-calculating and 
storing the staples across the entire lattice, and using these to make the links 
at the next level of fattening. Using this approach $\overline{\rm Fat7}$ requires 
less memory than HYP. In future work we hope to determine whether 
$\overline{\rm Fat7}$-blocked ensembles can produce results which are superior to HYP 
for certain calculations.

\section{ACKNOWLEDGEMENTS}
We wish to thank Anna Hasenfratz for discussing the relation of this
work to her own, and Derek Leinweber for bringing
refence~\cite{toomuch} to our attention.

\end{document}